\documentclass[manuscript]{aastex}

\usepackage{color}

\shorttitle{Magnetic reconnection detonation in supernova remnants}
\shortauthors{H. Zhang, Y. Gao, C. K. Law}

\begin{document}


\title{Magnetic reconnection detonation in supernova remnants}


\author{Horace Zhang\altaffilmark{1}}
\author{Yang Gao\altaffilmark{2, 3, *}}
\author{Chung K. Law\altaffilmark{2, 4}}
\affil{$^1$ Department of Physics, Princeton University, Princeton, New Jersey 08544, USA}
\affil{$^2$ Center for Combustion Energy, Tsinghua University, Beijing 100084, China}
\affil{$^3$ School of Physics and Astronomy, Sun Yat-Sen University, Zhuhai, Guangdong 519082, China}
\affil{$^4$ Department of Mechanical and Aerospace Engineering, Princeton University, Princeton, New Jersey 08544, USA}
\affil{$^*$ Email: gaoyang25@mail.sysu.edu.cn}







\begin{abstract}
As a key process that refreshes the interstellar medium,
  the dynamics and radiative properties of the supernova remnant (SNR) expansion front not only reflect the physical environment of the old interstellar medium (ISM) surrounding the supernova,
  but they also provide information about the refreshed ISM.
However the expansion dynamics of SNRs cannot be simply explained by the conventional law of spherical
  shock wave propagation;
   on the other hand, the high energy radiation requires an additional electron acceleration mechanism in the shock front beyond thermal collision.
We consider herein the detonation wave description of the SNR expansion,
  in which magnetic reconnection follows the shock front
  and transfers the SNR magnetic field energy to both fluid thermal energy
  and particle kinetic energy.
The structure of the magnetic reconnection detonation (MRD) is identified based on scaling analysis in this paper.
By applying the MRD description of the SNR expansion shock to the example of the Crab Nebula,
 this paper shows that the MRD description can explain both the accelerative expansion of the nebula as well as the origin of the luminous expanding shell.

\end{abstract}





\section{Introduction}

\begin{table}
\begin{center}
\begin{tabular}{ccccccc}
\hline
\hline
SNR  & Type & Age  & Expansion speed  &  \multicolumn{2}{c}{Expansion index $\eta$} & Reference \\ \cline{5-6}
         &      &  (yr) &  (km/s) & Radio  & X-ray     &  \\
\hline
Tycho    & Ia   & 430   & 3300    & 0.47   & 0.71, 0.33-0.65 \dag&   a-c    \\
Kepler   & Ia   & 390   & 4800    & 0.5    & 0.93, 0.35-0.8 \dag &   d-g    \\
Crab     & II   & 960   & 2300    & 1.26   & 1.04$^\ddag$        &   h      \\
\hline
\end{tabular}
\end{center}
 \caption{Expansions of three SNRs. $\dag$ From different observations. $\ddag$ Optical. (a) Strom (1982), (b) Hughes (2000), (c) Katsuda et al. (2010), (d) Dickel et al. (1988), (e) Hughes (1999), (f) Katsuda et al. (2008), (g) Vink (2008), (h) Bietenholz \& Nugent (2015).
}
\end{table}

For the expansion of the supernova remnants (SNRs),
  it is known that initially there exists an ejecta-dominated phase
  in which the freely expanding ejecta transfers their energy to the blast shock shell \citep{truelove1999}.
About several hundreds to a thousand years later the Sedov-Taylor phase becomes dominant, with the expansion index $\eta$
  (defined as $r \propto t^\eta$) being
  around $2/5$ in this shock wave \citep{sedov1993},
  or $3/5$ if an inhomogeneous surrounding environment is considered \citep{mckee1977}.
Although it is noted that the larger than $3/5$ expansion index may be theoretically explained by assuming radial density gradients
  in the ambient gas \citep{chevalier1982},
  under near-uniform ambient density the expansion index should always be smaller than 3/5 in the Sedov-Taylor phase.
Recent developments in radio and X-ray telescopes have enabled the direct measurement of SNR expansion velocities,
  showing that, for example, the expansion index $\eta$ of Tycho's SNR from radio observations falls between $2/5$ and $3/5$
  (Table 1, cf. \citet{strom1982}).
However, X-ray observations of the same SNR show an azimuthal anisotropy of expansion indices
  that gives $\eta$ values exceeding $3/5$ in the fast expanding directions \citep{hughes2000,katsuda2010,williams2016}.
Another SNR of SN Ia, Kepler, is also azimuthally anisotropic in its dynamics, with $\eta>3/5$ in some directions \citep{hughes1999,katsuda2008,vink2008}.
These observations suggest that both the Tycho and Kepler SNRs are likely evolving toward the Sedov-Taylor phase,
  while the larger than $3/5$ expansion index may hint at the existence of a local energetic process that speeds up the shock wave.

For SNRs with central pulsars, such as Crab nebula,
  the lost pulsar rotation energy can be transferred through the pulsar wind to support the expansion of the SNR.
This possibly leads to a larger-than-unity expansion index
  \citep[consistant with observations by][]{bietenholz1991,nugent1998,bietenholz2015}
  under the assumptions that the SNR is swept into a thin shell
  and that the ambient medium surrounding the pulsar bubble is expanding at a constant speed (i.e., a freely expanding SN ejecta).
This pulsar wind nebulae (PWNe) model, well established in \citet{kennel1984}, with the expansion of the PWNe - SN ejecta interface further studied
  in \citet{chevalier1992}, is considered to be the standard model of SNRs such as Crab in the sense that it explains both the radiation features and the
  dynamics of the nebula expansion very well.
However, there are still two uncertainties regarding the PWNe model for the Crab:
  (1) the expected `freely expanding envelope beyond the synchrotron nebula' is still not assuredly observed \citep{hester2008};
  and (2) in the PWNe model, the possibility that the relativistic pulsar wind caught up with and went through the SN ejecta in the first several decades after the SN explosion is not considered.
Recent works by \citet{yang2015,blondin2017} show the possibility that the Crab is the remnant of an underluminous SN in which
  the pulsar wind nebula breaks out the ejecta of SN, leaving most of the ejecta inside the observable nebula.
While if we consider such possibility to drop the assumption of a surrounding SN ejecta beyond the nebula
  (instead the observed nebula is expanding into a stationary ambient with a uniform density),
  the shock expansion index is $\sim 3/5$, being smaller than unity \citep[cf.][]{chevalier1984}.
Then the observations of the expansion index $\eta\gtrsim 1$ suggest additional sources of energy release associated with
  the SNR forward shock propagating into the interstellar medium (ISM),
  in which case the propagating front can be an accelerative expanding detonation wave \citep{gao2011}.

On the other hand,
  recent observations of $\gamma$-ray flares \citep{tavani2011,abdo2011} and long-term high energy emissions \citep{giordano2012,hess2015} from SNRs
  require an electron acceleration mechanism beyond thermal collision,
  and magnetic reconnection (MR) is a potential acceleration process \citep{yamaguchi2014,mochol2015}.
Follow-up observations of the $\gamma$-ray flares in Crab indicate that these high energy emissions happen close to the pulsar wind termination
  shock, i.e. about 0.1 pc apart from the pulsar \citep{schweizer2013}.
  As there are also unexpected high-energy emissions seen in the PWNe expansion shock at $\sim$ 2 pc \citep{hester2008},
  we consider the possible connection between shocks and high-energy radiations.
The high energy emissions of several GeV and above in the flares require fast, non-thermal electrons in synchrotron emission models.
Shock waves are usually considered to be the sites of such particle acceleration,
  and SNR shocks are deemed collisionless because the shock thickness is much shorter than the mean free path of electrons.
The magnetic field here then plays the role of speeding up the electrons,
  with the temporal and spatial evolution of magnetic fields creating strong electric fields by Faraday induction,
  and further incubating such high-energy electrons for high-energy emission through the synchrotron process.

In addition to accounting for the observed radiation, MR also transfers a significant part of the magnetic field energy
  to the thermal energy of the fluid \citep{yamada2014},
  which may serve as an additional energy source pushing forward the SNR,
  hence affecting the dynamics of its expansion.
Although it is demonstrated that magnetic field energy is only a small fraction of the overall SNR energy,
  MR can also affect the SNR dynamics locally and temporarily, as will be specified in this paper.
Recent simulation confirms the possibility of MR events in the strong shock downstream,
  where magnetic reconnection is induced by fluid turbulence downstream of the shock \citep{matsumoto2015}.
This kind of shock induced MR can occur in SNR, where the magnetic field inside the SNR is firstly amplified,
  and then MR is induced in the strong turbulent magnetized region downstream of the shock.
Such a delayed MR detonation is indirectly similar to the flame acceleration in obstructed channels reported recently \citep{bychkov2008}.
Consequently it is reasonable to consider the possibility that MR occurs in SNRs and as such explains the observed high-energy emissions as well as
  the $\eta\gtrsim 1$ expansion dynamics for the Crab and $\eta\gtrsim 3/5$ for the Tycho and Kepler SNRs.

\section{Magnetic reconnection detonation (MRD): formulations}

Considering MR occurring downstream of a hydrodynamic shock, without going into details about the MR process,
  the shock and the MR zone can be considered as an entire hydrodynamic transit front.
For simplification, we look into the case in which the upstream (ISM) magnetic field is negligible and
  the downstream (SNR) magnetic field is parallel to this transit front between the ISM and SNR.
Then we can employ the jump conditions across this transit front in one dimension \citep{draine1993}:
  \begin{eqnarray}
  \rho_1 v_1=\rho_2 v_2, \label{mass} \\
  \rho_1 v_1^2 +p_1 =\rho_2 v_2^2 +p_2 + \frac{B_2^2}{2\mu_0}, \label{momentum}\\
  \frac{1}{2}\rho_1 v_1^3 + \rho_1 h_1 v_1 = \frac{1}{2} \rho_2 v_2^3 + \rho_2 h_2 v_2, \label{energy}
  \end{eqnarray}
  where $\rho$, $v$, $p$, $B$, $h$ and $\mu_0$ are the density, velocity, pressure, magnetic field, enthalpy and magnetic conductivity in vacuum, respectively;
  and subscripts 1 and 2 denote the upstream and downstream variables, respectively;
  the upstream (ISM) magnetic field $B_1$ is assumed to be zero and hence does not show up in the formulation.

Based on the jump conditions we next derive the Rayleigh and Hugoniot relations for the hydrodynamic solution
  of this shock + MR transit front \citep[cf.][]{law2006}.
The Rayleigh relation can be expressed as
  \begin{eqnarray}
  (\hat{p}+\hat{p}_{\rm B})-1= - \gamma M_1^2(\hat{V}-1), \quad \rm{or} \label{layleigh 2} \\
  (\hat{p}+\hat{p}_{\rm B})-1= - \gamma M_2^2\frac{\hat{p}}{\hat{V}}(\hat{V}-1),  \label{layleigh 3}
  \end{eqnarray}
  where $V=1/\rho$ is the specific volume;
  and the dimensionless variables $\hat{p}=p_2/p_1$, $\hat{p}_{\rm B}=B_2^2/(2\mu_0 p_1)$, $\hat{V}=V_2/V_1$;
  and the Mach number $M=v/c_{\rm s}$ with
  $c_{\rm s}=\sqrt{\gamma p/\rho}$ the sound speed and $\gamma$ the polytropic index.
The Hugoniot relation has the form
  \begin{equation}
  [(\hat{p}+\hat{p}_{\rm B})+\frac{\gamma-1}{\gamma+1}](\hat{V}-\frac{\gamma-1}{\gamma+1})
  = \frac{4\gamma}{(\gamma+1)^2} + 2 \hat{q}_{\rm B} \frac{\gamma-1}{\gamma+1}, \label{hugoniot 2}
  \end{equation}
  where $\hat{q}_{\rm B}=q_{\rm B}(\rho_1/p_1)$ is the dimensionless magnetic field energy,
  with $q_{\rm B}=-(h_2-h_1)+c_{\rm p}(T_2-T_1)$ the specific energy released through MR within the transit layer and converted to thermal energy.
Here $c_{\rm p}$ is the constant specific heat, $T$ the temperature, and the ideal gas equation of state has been used.

It is readily seen that the Rayleigh (\ref{layleigh 2}, \ref{layleigh 3}) and Hugoniot (\ref{hugoniot 2}) relations for this transit front
  are the same as the relations in the conventional combustion waves \citep[cf. equs. (7.1.5) and (7.1.11) in][]{law2006},
  with the downstream pressure being $p_2'=p_2+p_{\rm B}$ and the energy release being $q_{\rm B}$.
So we can similarly find the Chapman-Jouguet (CJ) detonation solution for this shock + MR transit front, i.e.,
  \begin{equation}
  M_1^2=1+\frac{(\gamma^2-1)\hat{q}_{\rm B}}{\gamma}\left\{1+\left[1+\frac{2\gamma}{(\gamma^2-1)\hat{q}_{\rm B}}\right]^{1/2}\right\}. \label{CJ2}
  \end{equation}
The classical Zeld$\acute{o}$vich - von Neumann - D$\ddot{o}$ring (ZND) structure of detonation is thus also applicable to this transit system,
  where the upstream is firstly compressed by the leading shock to a compressed high temperature state within which exothermic reactions are ignited,
  pushing product fluids to the downstream.
This compressed transit layer between the upstream and downstream is called the Neumann layer,
  where MR occurs.
So the Neumann layer is connected to the upstream via the shock,
  and to the downstream via the energy release by assuming
  that the magnetic field energy $q_{\rm B}$ is totally converted to the thermal energy in the downstream.
The thickness of the Neumann layer is denoted as $l_{\rm ig}$, the ignition length of the exothermic reaction.
In the instant model, the exothermic reaction is MR, with the ignition process being the formation of turbulence and the
  deformation of magnetic field lines leading to reconnection.
More detailed formulation can be found in \citet{law2006},
 and an illustration of the ZND detonation structure (for Crab nebula as an example) can be found in Fig. 1 \citep[cf.][]{gao2011}.

\begin{figure}
\includegraphics[scale=0.5]{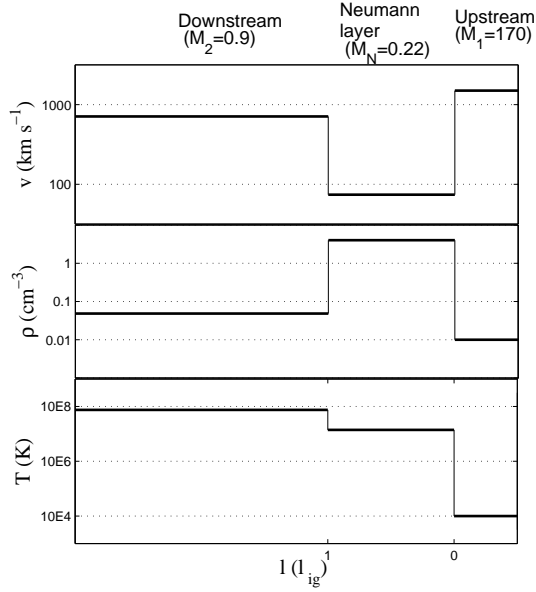}
\caption{ZND structure of the MR detonation for Crab nebula. The upstream is the ISM,
  the Neumann layer is the compressed ISM and the downstream is the nebula.
Typical density of the ISM, shock speed and nebula temperature are used in plotting the structure \citep{gao2011}.}
\end{figure}

We now proceed to apply this modified ZND detonation structure by inputting values appropriate to the Crab Nebula,
  and discuss its applicability to various SNRs.

\section{MRD scenario for Crab Nebula}

In an earlier work attempting to explain the accelerative expansion of Crab Nebula,
  the possibility of a detonation wave was proposed but without identifying the source of the exothermicity \citep{gao2011}.
Here we quantify the earlier suggestion that MR is indeed a suitable candidate for such an energy source.

The schematic illustration of this pulsar wind - detonation model is shown in Figure 2.
The most inner parts of pulsar - pulsar wind - pulsar wind termination shock (MHD shock) are the same as the standard PWNe model of \citet{kennel1984}
  where the pulsar spin down energy is transferred to the relativistic pulsar wind and then to the nebula via the MHD shock.
Beyond the MHD shock is the visible nebula, which is usually referred to as the `pulsar wind bubble' in literature \citep[e. g.][]{chevalier1992}.
In contrast to previous models, we consider the case that the relativistic pulsar wind forms early after the supernova explosion,
  and that the wind caught up and went through the whole SN ejecta to form an observed mixed wind-ejecta nebula region.
In such a situation, it can be easily inferred that the outer boundary of the nebula is the interface with the uniform, static interstellar medium (ISM).
Simulations by \citet{blondin2017} shows that it is possible for the ejecta to be totally penetrated by pulsar wind;
  here we consider the case that this happens shortly after the supernova explosion.
Assuming that a relativistic pulsar wind (with speed $\sim$ c the speed of light) begins 10 years after the explosion
  which throws ejecta at a maximal speed of $\sim 10^4$ ${\rm km}~{\rm s}^{-1}$,
  the pulsar wind catches up with the ejecta at a radius of $\sim 0.1$ pc for the Crab.
This is where the MHD shock forms.
From this time on, the downstream of the MHD shock carries the ejecta to further expand and interact with the ISM, forming the expansion detonation.

Then how can we understand the nature of nebula if it is a mixture of the pulsar wind and SN ejecta?
Recent numerical simulations on the propagation of pulsar wind - SN ejecta interface well resolve the development of the
  Rayleigh-Taylor structure and show how the pulsar wind blowout from a SNR \citep{blondin2017}.
In our assumption for the Crab nebula that the pulsar wind encounters SN ejecta $\sim$ 10 years after the explosion,
  this blowout can happen more rapidly so that the pulsar wind leads the expansion of the nebula for the remaining time of the nebula evolution.
On the other hand, when the pulsar wind encounters the SN ejecta,
  if the ejecta has already cooled down and is mostly in the recombination phase \citep{arnett1996,maurer2010}£¬
  one can assume that the SN ejecta is neutral before interacting with the pulsar wind.
Noticing that the MHD shock solution and nebula flow solution have been well established in \citet[cf., PWNe model of][]{kennel1984},
  and that in the solar wind - neutral ambient gas interaction the evolution of magnetic field follows almost the same trend as the PWNe \citep{holzer1972},
  we directly use the result regarding magnetic field evolution in these works.
It can be readily inferred that for fluid with small magnetic to inertial energy ratio, i.e., $\sigma=0.01$ in \citet{kennel1984}:
  the magnetic field is amplified by about 3 times in the nebula immediately downstream of the MHD shock,
  and further amplified in the downstream due to quasi-hydrodynamic effects.
The magnetic field then decreases following $1/r$ to about the same value of MHD shock upstream
  when it reaches the nebula outer boundary.
So, as the termination site of this non-neglectable magnetic field, the interface with the ISM
  (i.e., the expansion shock) deserves being revisited.

\begin{figure}
\includegraphics[scale=0.5]{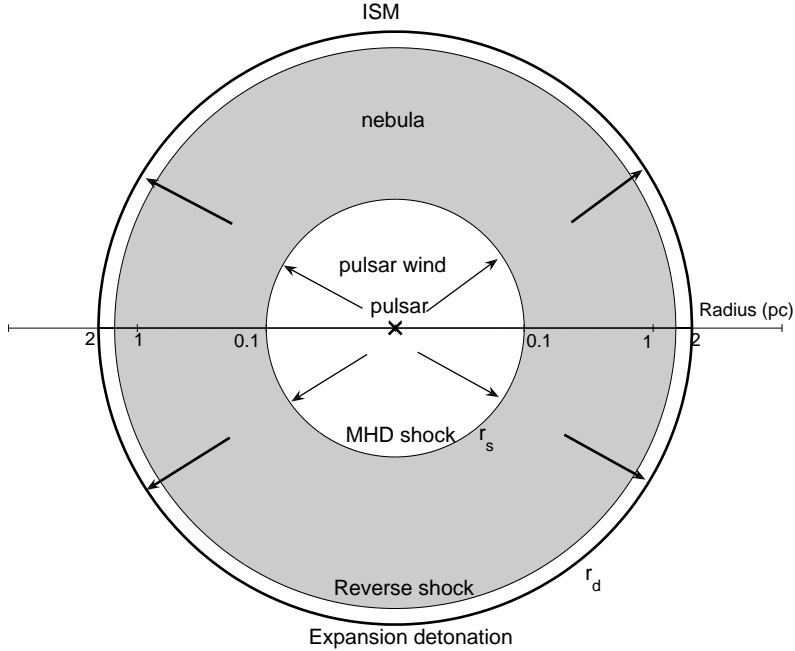}
\caption{Schematic structure for the Crab nebula.
The pulsar, pulsar termination shock (MHD shock) are the same as the standard PWNe model \citep{kennel1984}, with the region around the pulsar being
  the pulsar wind, terminated at the MHD shock.
Beyond the MHD shock is the nebula region, where the magnetic field is transported to the outer part.
In contrast to the PWNe model, the nebula region is a mixture of the pulsar wind and SN ejecta,
  as the pulsar wind penetrated the ejecta and breaks out to its outer edge, leading the expansion of the nebula.
At the outer edge of $r_{\rm d} = 2~ {\rm pc}$, a detonation wave forms as the nebula expands into the ISM,
  converting the magnetic field energy in the nebula to thermal and radiative energy through magnetic reconnection within the detonation.
Inside but close to the expansion detonation is the reverse shock, which amplifies the nebula magnetic field and transports it to
  the downstream of the expansion detonation.}
\end{figure}

\subsection{The feasibility of MR as the energy source in the detonation wave}

According to the acceleration of the nebula outer shell \citep{bietenholz1991},
  the energy density released in the Crab Nebula detonation reaction is estimated to be \citep{gao2011}
  \begin{equation}
  q_{\rm c}=\frac{ D_{\rm CJ}^2}{2(\gamma^2-1)}\approx5.9\times10^{12}{\rm J/mol}=61{\rm MeV/atom},\
  \end{equation}
  with $D_{\rm CJ}\sim1500$~km/s being the detonation speed \citep{hester2008} and $\gamma=1.1$ the polytropic index.
The polytropic index $\gamma=1.1$ adopted here is larger than the adiabatic index in the conventional PWNe, being 4/3 \citep{kennel1984}.
In the conventional PWNe, it has been demonstrated that the radiation loss of energy takes only $\sim 10 \%$ of the total nebula energy,
  thus the nebula is considered nearly adiabatic.
However, in the expansion detonation considered here, both the upstream near-isothermal ISM \citep[cf.][]{spaans2000}
  and the downstream adiabatic nebula are merged into
  the detonation Rankine-Hugoniot relations (\ref{layleigh 2}-\ref{hugoniot 2}), so a polytropic index between the two extremes is adopted.
We have estimated the energy release $q_{\rm c}$ by allowing the polytropic index to vary between 1.05 and 1.4,
  giving a $\sim$ 4 times variation of $q_{\rm c}$,
  which does not change the conclusion made later that MR energy is large enough to account for the detonation dynamics.

For comparison, the magnetic field energy (in the Neumann layer, immediately downstream of the shock)
  that can be released through MR is
  \begin{equation}
  P_{\rm B}=\frac{B_{\rm N}^2}{2\mu_0}=\frac{\rho_{\rm N}^2}{\rho_2^2}\frac{B_2^2}{2\mu_0},
  \end{equation}
  where $B_{\rm N}$, $\rho_{\rm N}$ and $B_2$, $\rho_{\rm 2}$ are the magnetic fields and densities
    of the Neumann layer and detonation downstream (cf. Fig. 3) regions respectively.
By converting the reaction energy $q_{\rm c}$ to an energy density $P_{\rm c}$ in the same units with $P_{\rm B}$,
  i.e., $P_{\rm c}= q_{\rm c} \rho_{N}/N_{\rm A}$, with $N_{\rm A}$ being the Avogadro number,
  we can readily estimate the ratio between the magnetic field energy and the \emph{required} reaction energy
  by taking the Crab magnetic field as $B_2\sim1~{\rm mG}$ \citep{hester2008,tavani2011,abdo2011}:
  \begin{equation}
  \frac{P_{\rm B}}{P_{\rm c}}=\frac{(\gamma^2-1)\rho_{\rm N} B_2^2 N_{\rm A}}{\mu_0 D_{\rm CJ}^2\rho_{\rm 2}^2}
  =\frac{\gamma^2 B_2^2 N_{\rm A}}{\mu_0 D_{\rm CJ}^2\rho_{\rm 1}}
  \approx 0.25.
  \label{energy ratio}
  \end{equation}
Here we adopted a local magnetic field $B_2$ which is larger than the average nebula value of $\sim300$ $\mu$G,
  because the magnetic field may be amplified through the reverse shock of the nebula expansion, as can be seen in Fig. 2.
Here we assume Chapman-Jouguet (CJ) detonation,
  that the upstream Mach number $M_{\rm 1}\gg1$,
  that the density of the Neumann state and its downstream
  are respectively $\rho_{\rm N}=\frac{\gamma+1}{\gamma-1}\rho_1$, and $\rho_2=\frac{\gamma+1}{\gamma}\rho_1$,
  and the upstream (ISM) density $\rho_1=0.01~{\rm cm}^{-3}$ for the Crab Nebula \citep{hester2008}.
Noted that as no `invisible' or `swept up' mass out of the nebula has been assumed in the current model, the upstream is assumed to be typical ISM.
Equation ({\ref{energy ratio}}) shows that the energy ratio depends on the strength of the magnetic field, the speed of the detonation wave,
  the density of the upstream ISM, and the polytropic index of the medium.
Furthermore, the variation of $B_2$, which is not an accurately measured parameter, strongly influences this ratio.
However, because this energy ratio is close to unity, MR is a reasonable candidate for
  the exothermic reaction in the detonation model if possible variations of physical variables in ({\ref{energy ratio}}) are considered.

It should be emphasized that in the above estimation, as well as in establishing the formulations in Section 2,
  the magnetic energy release is assumed to occur in the Neumann layer where the MR takes place.
Furthermore the magnetic field in the Neumann layer is related to the downstream (SNR) by the magnetic field frozen in subsonic flow
  $B_{\rm N}/\rho_{\rm N}=B_2/\rho_2$,
  with the supersonic upstream (ISM) magnetic field neglected (in the shock-front static framework).
In this sense the Neumann layer is the termination site of the SNR magnetic field,
   which is first amplified in the reverse shock and then further amplified in the contact discontinuity between the detonation downstream and the Neumann layer \citep[cf. Fig. 2 here and Fig. 47 in][]{lozinskaya1991}.
The existence of the Neumann layer (the density of which is also larger than the downstream SNR, cf. Fig. 1)
  is consistent with the theoretical sketch that involves the compressed parts ahead of the expanding SNR.


Although the comparison between the available magnetic field energy and the required detonation energy serves as the first estimate of the MRD model,
  one must further check whether MR occurs sufficiently fast to provide the required energy release.
Taking the Crab as an example,
  the ignition length\footnote{Ignition length of detonation is defined in the Neumann layer after the shock compression, indicating how long the flow propagates before exothermic reaction starts.} of the Crab Nebula detonation estimated from the critical radius of the detonation ignition is $l_{\rm ig}\sim1.0\times10^{-4}$~pc
  \citep{gao2011}.
This length, divided by the flow speed in the Neumann layer, $u_{\rm N}=74$~km/s, gives the ignition delay time of
  \begin{equation}
  \tau_{\rm ig}=4\times10^7~{\rm s}.
  \end{equation}
The typical time scale for MR in the Neumann layer can be estimated as \citep{ji2011}
  \begin{equation}
  \tau_{\rm rx}=\frac{l_{\rm ig}}{M_{\rm rx}v_{\rm A}},
  \end{equation}
  where $M_{\rm rx}$ is the reconnection rate with a value between 0.01-0.1 and
  \begin{equation}
  v_{\rm A}=\frac{B_{\rm N}}{\sqrt{\mu_0 m_{\rm H} \rho_{\rm N}}}=\frac{\gamma B_2}{\sqrt{(\gamma^2-1)\mu_0 m_{\rm H} \rho_1}}
  \end{equation}
  is the Alfv\'{e}n speed in the Neumann layer, where the SNR is assumed to be dominated by hydrogen atoms of mass $m_{\rm H}$.
Actually the SNR nebula is usually composed of heavier elements, with the average atomic mass being several times the mass of hydrogen.
This difference, as can be seen from eqns. (12) and (13), leads to only less than 2 times variation of the reconnection time,
  thus does not change the conclusion made here.
Assuming the reconnection rate $M_{\rm rx}=0.01$, and taking the values ($B_2\sim1~{\rm mG}$ and $\rho_1=0.01~{\rm cm}^{-3}$)
  of the Crab Nebula as used in estimating the energy ratio,
  the ratio between the reconnection time and the detonation ignition time is estimated as
  \begin{equation}
  \tau_{\rm rx}/\tau_{\rm ig}\approx0.15.
  \end{equation}
This ratio is less than unity, suggesting that the magnetic field within the Neumann layer has enough time to reconnect and
  transfer the magnetic field energy to the thermal energy.

\subsection{Structure, radiation and dynamics of MRD in Crab}

\begin{figure}
 \epsscale{1} \plotone{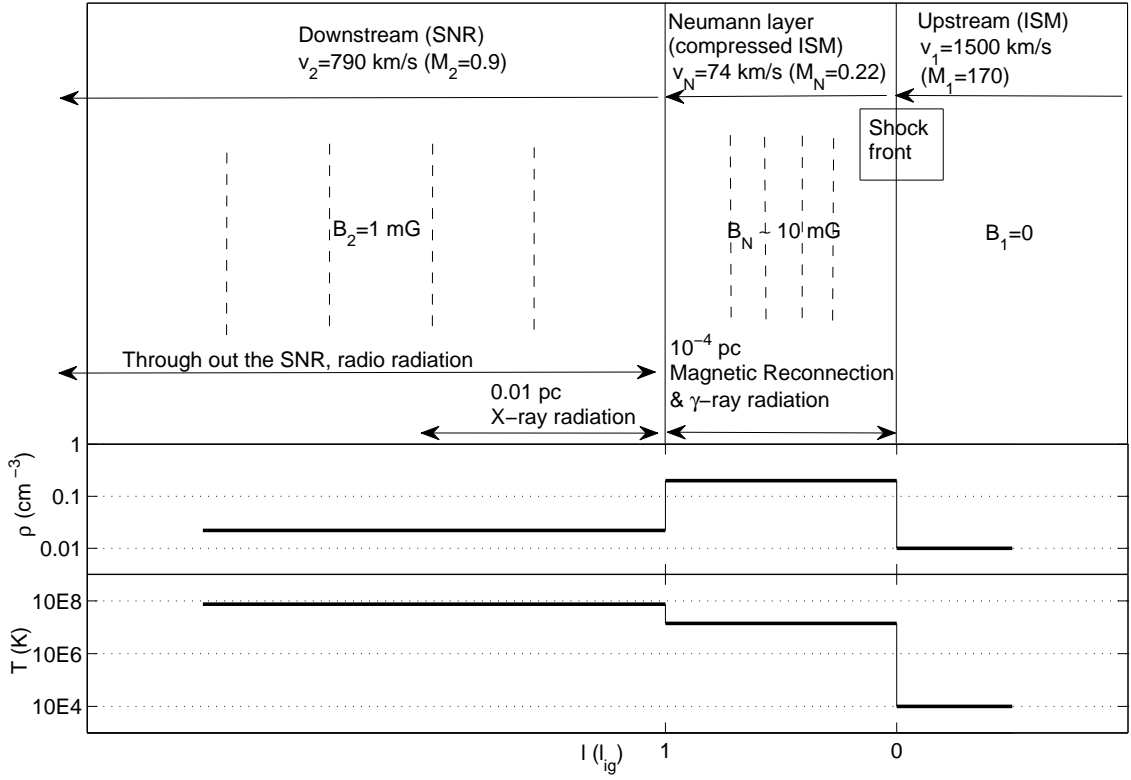}
\caption{Schematic of the MRD model for SNR, Crab Nebula as an example.
The upper panel shows the three-layer structure of the detonation front composed of the upstream ISM, the Neumann layer,
  and the downstream SNR, with the flow speed (in reference to the shock front) and local Mach number in each region indicated.
Here the upstream magnetic field is neglected, and the Neumann layer magnetic field is $\sim$ 10 times the SNR magnetic field,
  with the ratio equal to the density ratio when assuming that the magnetic field is frozen to the fluid.
The Neumann layer is where magnetic reconnection occurs to transfer the magnetic field energy to both the fluid thermal energy and
  to the kinetic energy of high energy non-thermal charged particles.
Following (and even within) the Neumann layer, non-thermal particles will first produce $\gamma$-ray and X-ray radiation,
  while radio and optical emissions with lower energy spread over the length of SNR of a few parsecs.
The two lower panels show the variation of density and temperature in the three layers as a reference (cf. Fig. 1).}
\end{figure}

We next consider the size of the radiation zone within the MRD structure.
For the case of Crab, the synchrotron radiation loss time can be estimated by using $\tau_{\rm loss}\approx(8\times10^8~{\rm s})B^{-2}\Gamma^{-1}$
  \citep{tavani2011},
  where $B=B_2$ in Gauss is the SNR magnetic field within which radiation takes place, and $\Gamma$ is the Lorentz factor of electrons.
Then the radiation time for typical high-energy synchrotron emissions from electrons with $\Gamma=2\times10^6$,
  corresponding to critical photon energy of $\sim {\rm keV}$,
  is $\tau_{\rm lossH}\approx 4\times10^8~{\rm s}\sim10~{\rm yr}$.
This value, multiplied by the typical sound speed within the Crab SNR, $a\approx 800~{\rm km/s}$ \citep{gao2011},
  yields an estimate for the thickness of the high-energy radiation zone, $\sim0.01~{\rm pc}$ (Fig. 3).
For electrons with even larger Lorentz factor, i.e., $\Gamma\sim10^9$, corresponding to GeV synchrotron photons,
   the emission zone is much thinner ($\sim 10^{-4}$~pc),
   indicating that such extreme-high-energy radiation occurs within the Neumann layer, immediately after the reconnection takes place.
Using the same equations, the thickness of the radiation zone for low-energy radio emissions, corresponding to $\Gamma\sim 1$,
  is $\sim10^4~{\rm pc}$.
This is much larger than the size of the SNR,
  meaning that the radio radiation occurs throughout the SNR (Fig. 3).

According to simulations by \citet{cerutti2013}, the high energy electrons can be generated in MR,
  leading to synchrotron radiation above the conventional limit of 160 MeV \citep{dejager1996}.
However when the electron gyroradius exceeds the thickness of reconnection zone (Neumann layer),
  fast particles will escape to the inner nebula region where the magnetic field is smaller.
This forms another cutoff for the high energy radiation.
The radiation features discussed above are those related to the expansion detonation, located at the outer shell of the SNR.
If we look at the inner part of the nebula,
  both the magnetic field and density of relativistic particles are higher compared to the outer part of nebula,
  so most of the high energy emissions should still locate at the inner nebula.
Just in addition to this picture, our model predicts that synchrotron emissions are amplified in the outer shell,
  with transitory high and low energy emissions (e.g., flares) caused by MR activities observed at times.
As MR can also occur in the downstream of the MHD shock (cf. Fig. 2), where both the particle energy and magnetic field are higher,
  observed high energy $\gamma$-ray flares can be induced there as well \citep{schweizer2013}.
Thus, this model radiation feature of Crab nebula is consistent with current observations.

Based on the above scaling analysis of energy, reconnection time and radiation time, the proposed MRD model has the following features:
(1) There is a compressed Neumann layer with thickness $\sim 10^{-4}~{\rm pc}$, following the SNR forward shock front that separates
  the upstream ISM and the downstream SNR.
Within this layer, the fluid density is amplified to about 20 times that of the upstream ISM,
  and the magnetic field there is $\sim$10 times that in the downstream SNR.\footnote{It is also noted that streaming instability of the cosmic ray current in crossing the SNR shock front may amplify the upstream magnetic field to
  $\sim$ 100 times or even larger in shocks with Mach number of several hundreds \citep{riquelme2009,caprioli2014},
  which may be responsible for the magnetic field amplification of Tycho and Kepler SNRs.}
Within this layer, the magnetic field is highly turbulent and MR occurs
  transferring the magnetic field energy to both fluid thermal energy and non-thermal kinetic energy of charged particles.
(2) The additional thermal energy released through MR pushes ahead the front of the forward shock,
  forming the detonation wave which explains the close-unity expansion index for the Crab SNR.
(3) The non-thermal particles accelerated through MR radiate their kinetic energy in the downstream SNR through the synchrotron process.
  The typical thickness of the high-energy X-ray ($\sim$keV) emission zone is $\sim 0.01~{\rm pc}$
  while low-energy radio emission occurs throughout the entire SNR.

\section{Further considerations}

It is noted that the detonation mechanism of SNR propagation does not exclude the \citet{chevalier1982,chevalier1984} model,
  which additionally assumes density variation in the radial direction
  and as such may also contribute to the dynamic evolution of the SNR, in addition to the MRD effect.
So the MRD scenario presented here is one of the several possible reasons that the SNRs expand faster than
  the prediction of the Sedov solution or the conventional pulsar wind solution propagating into a stationary uniform ambient gas.
The feasibility of the MRD model requires further observational tests of its radiation properties.
For pulsar wind SNRs,
  it is natural to define the MRD expansion epoch as an SNR evolution phase after the initial blast wave expansion.
Comparison of the magnetic field energy with the exothermal energy required in the detonation wave model, i.e., equ. (10), shows that
  SNRs with relatively high magnetic field ($\gtrsim 1~{\rm mG}$, with 1 mG being a flexible rather than strict value)
  and low expansion speeds ($\sim1000~{\rm km/s}$) such as the Crab Nebula
  can have the MRD process account for the global expansion of the nebula.

For Type Ia SNRs, the magnetic field is usually low ($\lesssim 1~{\rm mG}$) and the expansion speeds is high
  ($\sim 3000~{\rm km/s}$) such as in the Tycho and Kepler,
  the MRD scenario can only take place after the SNR expansion further slows down and becomes comparable to the detonation speed
  inferred from the magnetic field energy in the shock downstream.
Even when such condition is achieved, as there is no continuous feed up of magnetic field to the nebula,
  the MRD is only a transitory epoch that exists in those slowly expanding, strongly magnetized SNRs.
Additionally, in young shell-like SNRs the magnetic field is usually in the radial direction up to the outer edge \citep{dubner2015},
  and magnetic field amplification cannot take place during the MHD shock and the reverse shock compressions.
However, MR may still occur due to the turbulence induced by the shocks.
Then in this case the speed-up of supernova remnant should not be expected,
  but the high energy emissions in the shock downstream still exist.

For both pulsar wind and type Ia SNRs, by considering the azimuthal variation of the magnetic field,
  it is also possible that in the direction where the local magnetic field is several ${\rm mG}$ or higher,
  the detonation speed of MRD is comparable to the SNR expansion speed,
  making the expansion index in this direction close to unity.
Validation of this prediction requires observations of both the high-energy emissions and the high-resolution expansion dynamics
  of SNRs to check the existence of local MRDs and their connections with flares.

\acknowledgments
We appreciate the discussions with Prof. Hantao Ji and Prof. Amitava Bhattacharjee on the magnetic reconnection.
This work was supported by the Center for Combustion Energy at Tsinghua University and by the National Science Foundation of China grant 51206088. Y.G. acknowledges support by the Princeton Plasma Physics Laboratory and the Tsinghua-Santander Program for young faculty performing research abroad.

\clearpage




\clearpage


\begin{thebibliography}{}


\bibitem[Abdo et al. (2011)]{abdo2011}
Abdo, A. A., Ackermann, M., Ajello, M., et al. 2011, Science, 311, 739

\bibitem[Arnett (1996)]{arnett1996}
Arnett, D. 1996, Supernovae and Nucleosynthesis (Princeton University Press, Princeton, New Jersey)

\bibitem[Bietenholz et al. (1991)]{bietenholz1991}
Bietenholz, M. F., Kronberg, P. P., Hogg, D. E. \& Wilson, A. S. 1991, ApJ, 373, L59

\bibitem[Bietenholz \& Nugent (2015)]{bietenholz2015}
Bietenholz, M. F. \& Nugent, R. L. 2015, MNRAS, 454, 2416

\bibitem[Blondin \& Chevalier (2017)]{blondin2017}
Blondin, J. M. \& Chevalier, R. A. 2015, ApJ, 845, 139

\bibitem[Bychkov et al. (2008)]{bychkov2008}
Bychkov, V., Valiev,D., \& Eriksson, L.-E. 2008, Phys. Rev. Lett. 101, 164501


\bibitem[Caprioli \& Spitkovsky (2014)]{caprioli2014}
Caprioli, D. \& Spitkovsky, A. 2014, ApJ, 794, 46

\bibitem[Cerutti et al. (2012)]{cerutti2012}
Cerutti, B., Uzdensky, D. A. \& Begelman M. C. 2012 ApJ, 746, 148

\bibitem[Cerutti et al.(2013)]{cerutti2013}
Cerutti, B., Werner, G. R., Uzdensky, D. A. \& Begelman, K. C. 2013, \apj, 770, 147




\bibitem[Chevalier (1982)]{chevalier1982}
Chevalier, R. A. 1982, ApJ, 258, 790

\bibitem[Chevalier (1984)]{chevalier1984}
Chevalier, R. A. 1984, ApJ, 280, 797

\bibitem[Chevalier \& Fransson (1992)]{chevalier1992}
Chevalier, R. A. \& Fransson C. 1992, ApJ, 395, 540

\bibitem[De Jager et al.(1996)]{dejager1996}
De Jager, O. C., Harding, A. K., Michelson, P. F. et al., 1996, \apj, 457, 253

\bibitem[Dickel et al. (1988)]{dickel1998}
Dickel, J. R., Sault, R., Arendt, R. G., Matsui, Y. \& Korista, K. T. 1988, ApJ, 330, 254

\bibitem[Draine \& McKee (1993)]{draine1993}
Draine, B. T. \& McKee C. F. 1993, ARAA, 31, 373
%

\bibitem[Dubner \& Giacani (2015)]{dubner2015}
Dubner, G. \& Giacani, E. 2015, A\&ARv, 23, 3

%

\bibitem[Gao \& Law (2011)]{gao2011}
Gao, Y. \& Law, C. K. 2011, Phys. Rev. Lett., 107, 171102

\bibitem[Giordano et al. (2012)]{giordano2012}
Giordano, F., Naumann-Godo, M, Ballet, J. et al. 2012, ApJ, 744, L2

\bibitem[Hester (2008)]{hester2008}
Hester, J. J. 2008, ARAA, 46, 125

\bibitem[Holzer (1972)]{holzer1972}
Holzer, T. E. 1972, Journal of Geophysical Research, 77, No. 28, 5407

\bibitem[Hughes (1999)]{hughes1999}
Hughes, J. P. 1999, ApJ, 527, 298

\bibitem[Hughes (2000)]{hughes2000}
Hughes, J. P. 2000, ApJ, 545, L53

\bibitem[Ji \& Daughton (2011)]{ji2011}
Ji, H. \& Daughton, W. 2011, Phys. Plasmas, 18, 111207

\bibitem[Katsuda et al. (2008)]{katsuda2008}
Katsuda, S., Tsunemi, H., Uchida, H. \& Kimura, M. 2008, ApJ, 689, 225


\bibitem[Katsuda et al. (2010)]{katsuda2010}
Katsuda, S., Petre, R., Hughes, J. P. et al. 2010, ApJ, 709, 1387

\bibitem[Kennel \& Coroniti (1984)]{kennel1984}
Kennel, C. F. \& Coroniti, F. V. 1984, ApJ, 283, 694

\bibitem[Law (2006)]{law2006}
Law, C. K. 2006, Combustion Physics (Cambridge University Press, New York)

\bibitem[Lozinskaya (1991)]{lozinskaya1991}
Lozinskaya, T. A. 1991, Supernovae and Stellar Wind in the Interstellar Medium
(American Institute of Physics, New York)


\bibitem[Matsumoto et al. (2015)]{matsumoto2015}
Matsumoto, Y., Amano, T., Kato, T. N. \& Hoshino M. 2015, Science, 347, 974

\bibitem[Maurer \& Mazzali (2010)]{maurer2010}
Maurer, I. \& Mazzali, P. A. 2010, MNRAS, 408, 947

\bibitem[McKee \& Ostriker (1977)]{mckee1977}
McKee, C. F. \& Ostriker, J. P. 1977, ApJ, 218, 148

\bibitem[Mochol \& P\`{e}tri (2015)]{mochol2015}
Mochol, I. \& P\`{e}tri, J. 2015, MNRAS, 449, 51

\bibitem[Nugent (1998)]{nugent1998}
Nugent, R. L. 1998, PASP, 110, 831

\bibitem[Riquelme \& Spitkovsky (2009)]{riquelme2009}
Riquelme, M. A. \& Spitkovsky, A. 2009, ApJ, 694, 626

\bibitem[Sedov (1993)]{sedov1993}
Sedov, L. I. 1993, Similarity and Dimension Methods in Mechanics (10th ed, CRC Press, London)

\bibitem[Strom et al. (1982)]{strom1982}
Strom, R. G., Goss, W. M. \& Shaver, P. A. 1982, MNRAS 200, 473




%

\bibitem[Spaans \& Silk (2000)]{spaans2000}
Spaans, M. \& Silk, J. 2000, ApJ 538, 115

\bibitem[Schweizer et al. (2013)]{schweizer2013}
Schweizer, T., Bucciantini, N., Idec, W. et al. 2013, MNRAS 433, 3325

\bibitem[Tavani et al. (2011)]{tavani2011}
Tavani, M., Bulgarelli, A., Vittorini, V. et al. 2011, Science, 331, 736

\bibitem[Truelove \& McKee (1999)]{truelove1999}
Truelove, J. K. \& McKee, C. F 1999, ApJS, 120, 299

\bibitem[HESS (2015)]{hess2015}
The H.E.S.S. Collaboration 2015, Science, 347, 46

\bibitem[Vink (2008)]{vink2008}
Vink, J. ApJ, 2008, 689, 231


\bibitem[Williams et al. (2016)]{williams2016}
Williams, B. J., Chomiuk, L., Hewitt, J. W. et al. 2016, ApJ, 823, L32

\bibitem[Yamada et al. (2014)]{yamada2014}
Yamada, M., Yoo, J., Jara-Almonte, J., Ji, H.-T. Kulsrud, R. M. \& Myers, C. E. 2014, Nature Commun., 5, 4774

\bibitem[Yamaguchi et al. (2014)]{yamaguchi2014}
Yamaguchi, H., Eriksen, K. A., Badenes, C. et al. 2014, ApJ, 780, 136

\bibitem[Yang \& Chevalier (2015)]{yang2015}
Yang, H. \& Chevalier, R. A. 2017, ApJ, 806, 153









\end{thebibliography}
\end{document}